\input{aipcheck}

\def\dd{{\rm d}}
\def\p{I\!\!P}

\def\lr{\left( }
\def\rr{\right) }
\def\le{\left[ }
\def\re{\right] }

\def\beq{\begin{equation}}
\def\eeq{\end{equation}}
\def\bea{\begin{eqnarray}}
\def\eea{\end{eqnarray}}

\documentclass[
    ,final            
  ]
  {aipproc}

\layoutstyle{6x9}

\begin{document}

\title{Diffractive Dijet Photoproduction}



\author{Michael Klasen}{address={Institute for Nuclear Theory, University of
 Washington, Box 351550, Seattle, WA 98195, USA {\em and} \\
 Laboratoire de Physique Subatomique et de Cosmologie, Universit\'e
 Joseph Fourier/CNRS-IN2P3, 53 Avenue des Martyrs, 38026 Grenoble, France}}

\author{Gustav Kramer}{address={II.\ Institut f\"ur Theoretische Physik,
 Universit\"at Hamburg, Luruper Chaussee 149, 22761 Hamburg, Germany}}

\begin{abstract}
We have calculated diffractive dijet production in deep-inelastic scattering
(DIS) at low-$Q^2$ and next-to-leading order (NLO) of perturbative QCD,
including contributions from direct and resolved photons. We study how the
cross section depends on the factorization scheme and scale $M_{\gamma}$ at
the virtual photon vertex for the occurance of factorization breaking. The
strong $M_{\gamma}$-dependence, which is present when only the resolved
cross section is suppressed, is tamed by intodrucing the suppression also in
the initial-state NLO correction of the direct part.
\end{abstract}

\maketitle

\vspace*{-11cm}
\noindent LPSC 05-055 \\
INT-ACK 05-039 \\
hep-ph/0507014
\vspace*{  9cm}


\section{Introduction}
\label{sec:1}

From a perturbative QCD (pQCD) point of view, the central question for hard
diffractive scattering events, characterized by a large rapidity gap devoid
of particles in high-energy collisions, is whether they can be factorized
into non-perturbative diffractive parton density functions (PDFs) of a
colorless object ({\it e.g.} a pomeron) and perturbative partonic cross
sections. This concept, based on a long-standing proposal by Ingelman and
Schlein \cite{Ingelman:1984ns}, is believed to hold for the scattering of
point-like electromagnetic probes off a hadronic target, such as
deep-inelastic scattering (DIS) or {\it direct} photoproduction
\cite{Collins:1997sr}, whereas it has been shown to fail for purely hadronic
collisions \cite{Collins:1997sr,Affolder:2000vb}.

Factorization is thus expected to fail also in {\it resolved}
photoproduction, where the photon first dissolves into partonic
constituents, before these scatter off the hadronic target. The separation
of these two types of photoproduction events is, however, a leading order
(LO) concept. At next-to-leading order (NLO) of pQCD, they are closely
connected by an initial state singularity originating from the splitting
$\gamma\to q\bar{q}$ \cite{Klasen:2002xb}, which may play a crucial role in
the way factorization breaks down in diffractive photoproduction
\cite{Klasen:2005dq}. Factorization breaking effects are therefore expected
to show up first in observables that distinguish between direct and resolved
photoproduction, such as distributions in the longitudinal momentum fraction
$x_\gamma$ of partons in the photon \cite{Klasen:2004tz}, the photon
virtuality $Q^2$ \cite{Klasen:2004ct}, or the dependence of the predicted
cross sections on the factorization scale $M_\gamma$ \cite{Klasen:2005dq}. 
It is clear that this $M_\gamma$-dependence is unphysical and must be
remedied also for the case of factorization breaking of the resolved part of
the cross section. A proposal how to achieve this has been worked out in our
previous work \cite{Klasen:2005dq} and will be described in the next
Section. For demonstrative purposes we restrict ourselves to low-$Q^2$ DIS
using the kinematic framework of our earlier publication
\cite{Klasen:2004ct}.

Three groups have recently tried to extract diffractive parton densities
from inclusive diffractive DIS data at DESY HERA, treating the Pomeron
either as a hadronic object within Regge factorization 
\cite{h1ichep02,Chekanov:2004hy}
\begin{equation}
 f_a^D(x,Q^2;x_{\p},t) = f_{\p/p}(x_{\p},t) f_{a/\p}(\beta=x/x_{\p},Q^2)
\end{equation}
or perturbative QCD \cite{Martin:2004xw}, but so far only the H1 set has
been tested for factorization in a different scattering environment,
{\it i.e.} in photoproduction of dijets at HERA \cite{Mozer}.

\section{Factorization Scheme and Scale Dependence}
\label{sec:2}

A factorization scheme for virtual photoproduction has been defined and the
full NLO corrections for inclusive dijet production have been calculated in
\cite{Klasen:1997jm}. They have been implemented in the NLO Monte Carlo
program JET\-VIP \cite{Potter:1999gg}. We have adapted this NLO framework to
diffractive dijet production. According to \cite{Klasen:1997jm}, the
subtraction term, which is absorbed into the PDFs of the virtual photon
$f_{a/\gamma}(x_\gamma,M_{\gamma})$, is of the form as given in
\cite{Klasen:2005dq}. The main term is proportional to
$\ln(M_{\gamma}^2/Q^2)$ times the splitting function
\beq
 P_{q_i \leftarrow \gamma}(z) = 2 N_c Q_i^2 \frac{z^2+(1-z)^2}{2},
 \label{eq:2}
\eeq
where $z=p_1p_2/p_0q \in [x;1]$ and $Q_i$ is the fractional charge of the
quark $q_i$. $p_1$ and $p_2$ are the momenta of the two outgoing jets, and
$p_0$ and $q$ are the momenta of the ingoing parton and virtual photon,
respectively. Since $Q^2=-q^2 \ll M_{\gamma}^2$, the subtraction term is
large and is therefore resummed by the DGLAP evolution equations for the
virtual photon PDFs. After this subtraction, the finite term
$M(Q^2)_{\overline{\rm MS}}$, which remains in the matrix element for the
NLO correction to the direct process \cite{Klasen:1997jm}, has the same 
$M_{\gamma}$-dependence as the subtraction term, {\it i.e.} $\ln M_{\gamma}$
is multiplied with the same factor. As already mentioned, this yields the
$M_{\gamma}$-dependence before the evolution is turned on. In the usual
non-diffractive dijet photoproduction these two $M_{\gamma}$-dependences
cancel, when the NLO correction to the direct part is added to the LO
resolved  cross section \cite{BKS}. Then it is obvious that the approximate
$M_{\gamma}$-independence is destroyed, if the resolved cross section is
multiplied by a suppression factor $R$ to account for the factorization
breaking in the experimental data. To remedy this deficiency, we propose to
multiply the $\ln M_{\gamma}$-dependent term in $M(Q^2)_{\overline{\rm MS}}$
with the same suppression factor as the resolved cross section. This is done
in the following way: We split $M(Q^2)_{\overline{\rm MS}}$ into two terms
using the scale $p_T^{*}$ in such a way that the term containing the slicing
parameter $y_s$, which was used to separate the initial-state singular
contribution, remains unsuppressed. In particular, we replace the finite
term  after the subtraction by
\bea
 M(Q^2,R)_{\overline{\rm MS}} &=& \le-\frac{1}{2N_c} P_{q_i\leftarrow
 \gamma}(z)\ln\lr\frac{M_{\gamma}^2 z}{p_T^{*2}(1-z)}\rr+{Q_i^2\over2} \re R
 \nonumber \\
 && \ -\frac{1}{2N_c} P_{q_i\leftarrow\gamma}(z)
 \ln\lr\frac{p_T^{*2}}{zQ^2+y_s s}\rr,\label{eq:4}
\eea
where $R$ is the suppression factor. This expression coincides with the
finite term after subtraction (see Ref.\ \cite{Klasen:2005dq}) for $R=1$, as
it should, and leaves the second term in Eq.\ (\ref{eq:4}) unsuppressed. In
Eq.\ (\ref{eq:4}) we have suppressed in addition to $\ln(M_{\gamma}^2/
p_T^{*2})$ also the $z$-dependent term $\ln (z/(1-z))$, which is specific to
the $\overline{\rm MS}$ subtraction scheme as defined in
\cite{Klasen:1997jm}. The second term in Eq.\ (\ref{eq:4}) must be left in
its original form, {\it i.e.} being unsuppressed, in order to achieve the
cancellation of the slicing parameter ($y_s$) dependence of the complete NLO
correction in the limit of very small $Q^2$ or equivalently very large $s$.
It is clear that the suppression of this part of the NLO correction
to the direct cross section will change the full cross section only very
little as long as we choose $M_{\gamma} \simeq p_T^{*}$. The first term in
Eq.\ (\ref{eq:4}), which has the suppression factor $R$, will be denoted by
${\rm DIR}_{\rm IS}$ in the following.

To study the left-over $M_{\gamma}$-dependence of the physical cross
section, we have calculated the diffractive dijet cross section with the
same kinematic constraints as in the H1 experiment \cite{Schatzel:2004be}. 
Jets are defined by the CDF cone algorithm with jet radius equal to one and
asymmetric cuts for the transverse momenta of the two jets required for
infrared stable comparisons with the NLO calculations \cite{Klasen:1995xe}.
The original H1 analysis actually used a symmetric cut of 4 GeV on the
transverse momenta of both jets \cite{Adloff:2000qi}. The data have,
however, been reanalyzed for asymmetric cuts \cite{Schatzel:2004be}. 

For the NLO resolved virtual photon predictions, we have used the PDFs SaS1D
\cite{Schuler:1996fc} and transformed them from the DIS$_{\gamma}$ to the 
$\overline{\rm MS}$ scheme as in Ref.\ \cite{Klasen:1997jm}. If not stated
otherwise, the renormalization and factorization scales at the pomeron and
the photon vertex are equal and fixed to $p_T^{*} = p_{T,jet1}^{*}$. We
include four flavors, {\it i.e.} $n_f=4$ in the formula for $\alpha_s$ and
in the PDFs of the pomeron and the photon. With these assumptions we have
calculated the same cross section as in our previous work
\cite{Klasen:2004ct}. First we investigated how the cross section
$\dd\sigma/\dd Q^2$ depends on the factorization scheme of the PDFs for the
virtual photon, {\it i.e.} $\dd\sigma/\dd Q^2$ is calculated for the choice
SaS1D and SaS1M. Here $\dd\sigma/\dd Q^2$ is the full cross section (sum of
direct and resolved) integrated over the momentum and rapidity ranges as in
the H1 analysis. The results, shown in Fig.\ 2 of Ref.\ \cite{Klasen:2005dq}
demonstrate that the choice of the factorization scheme
of the virtual photon PDFs has negligible influence on $\dd\sigma/\dd Q^2$
for all considered $Q^2$. The predictions agree reasonably well with the
preliminary H1 data \cite{Schatzel:2004be}. 

\begin{figure}
 \centering
 \includegraphics[width=.49\textwidth]{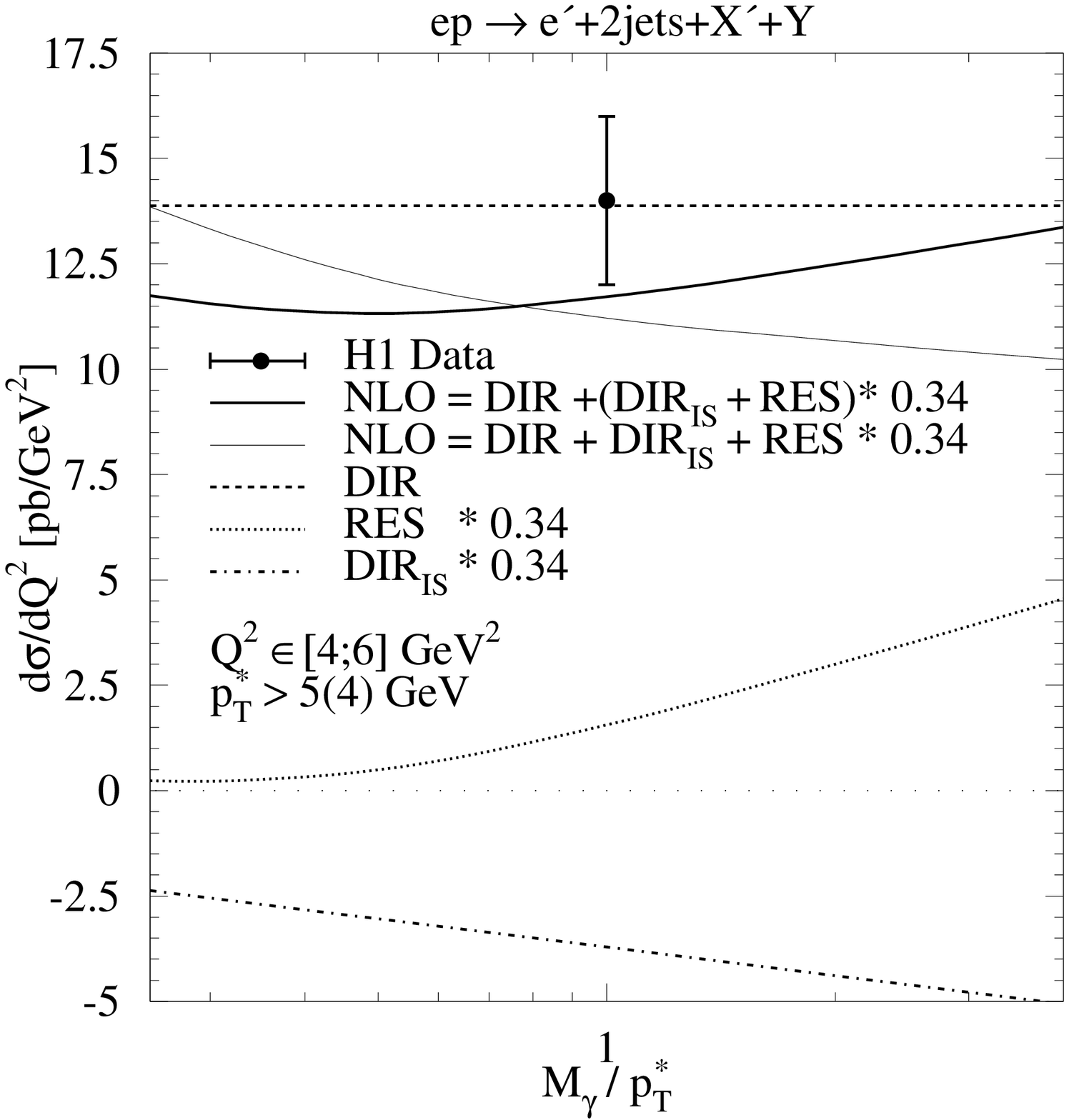}
 \includegraphics[width=.49\textwidth]{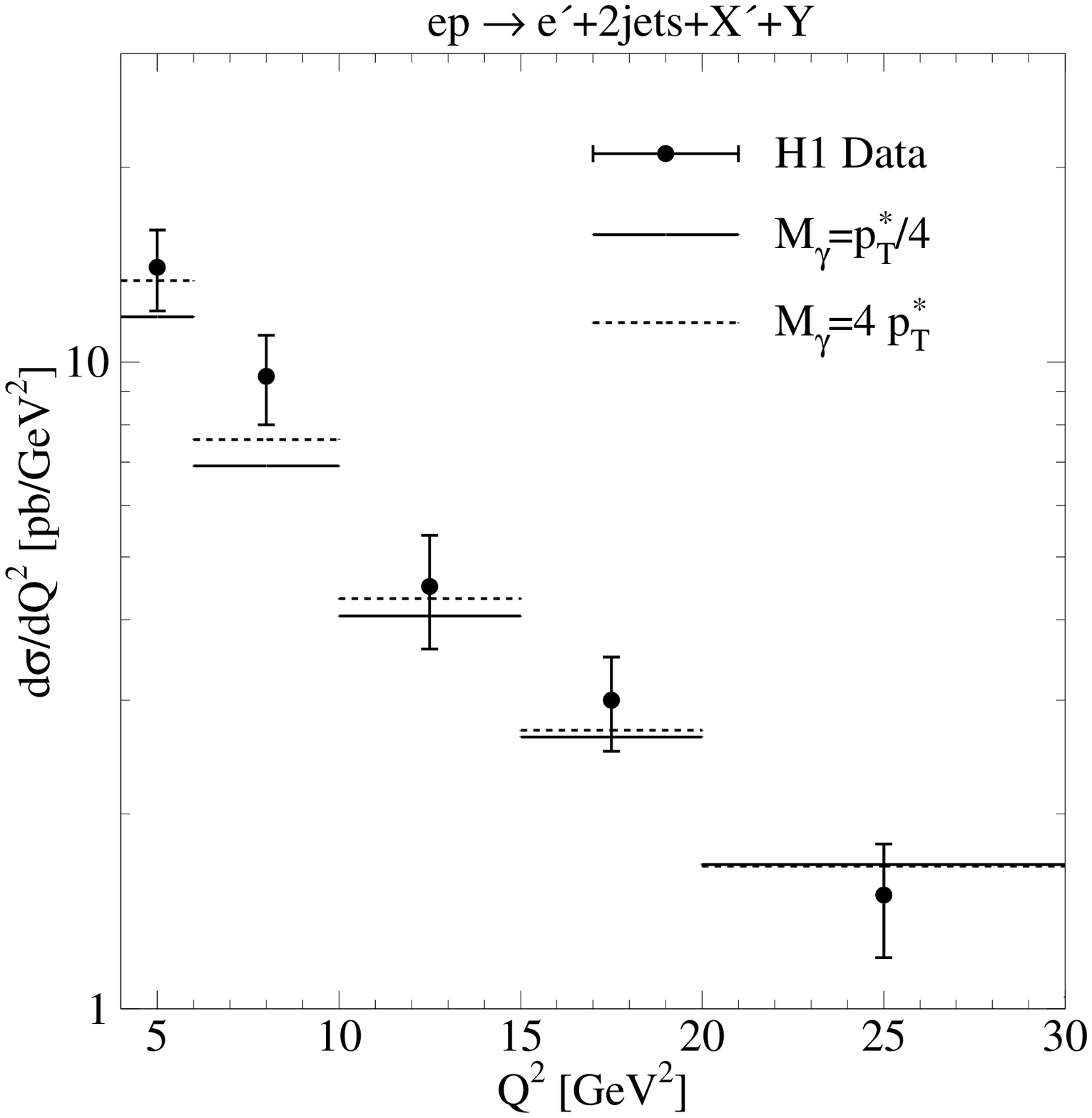}
 \caption{\label{fig:1}Left: Photon factorization scale dependence of
 resolved and direct contributions to $\dd\sigma/\dd Q^2$ together with
 their weighted sums for (i) suppression of the resolved cross section and
 for (ii) additional suppression of DIR$_{\rm IS}$, using SaS1D virtual
 photon PDFs \cite{Schuler:1996fc}. Right: $Q^2$-dependence of the dijet
 cross section for $M_{\gamma}= p^{*}_T/4$ (full) and $M_{\gamma}=4p^{*}_T$
 (dashed) and comparison with preliminary H1 data using SaS1D virtual photon
 PDFs \cite{Schuler:1996fc}.}
\end{figure}

We now turn to the $M_{\gamma}$-dependence of the cross section with a
suppression factor for DIR$_{\rm IS}$, which is the main part of this
Report. To show this dependence for the two suppression mechanisms, (i)
suppression of the resolved cross section only and (ii) additional
suppression of the DIR$_{\rm IS}$ term as defined in Eq.\ (\ref{eq:4}) in
the NLO correction of the direct cross section, we consider $\dd\sigma/
\dd Q^2$ for the lowest $Q^2$-bin, $Q^2\in [4,6]$ GeV$^2$. In the left part
of Fig.\ \ref{fig:1}, this cross section
%
%
is plotted as a function of $\xi=M_{\gamma}/p_T^{*}$ in the range $\xi\in
[0.25;4]$ for the cases (i) (light full curve) and (ii) (full curve). We see
that the cross section for case (i) has an appreciable $\xi$-dependence in
the considered $\xi$ range of the order of $40\%$, which is caused by the
suppression of the resolved contribution only. With the additional
suppression of the DIR$_{\rm IS}$ term in the direct NLO correction, the
$\xi$-dependence of $\dd\sigma/\dd Q^2$ is reduced to approximately less
than $20\%$, if we compare the maximal and the minimal value of $\dd\sigma/
\dd Q^2$ in the considered $\xi $ range. The remaining $\xi $-dependence is
caused by the NLO corrections to the suppressed resolved cross section and
the evolution of the virtual photon PDFs. How the compensation of the
$M_{\gamma}$-dependence between the suppressed resolved contribution and the
suppressed direct NLO term works in detail is exhibited by the dotted and
dashed-dotted curves in Fig.\ \ref{fig:1} (left). The suppressed resolved
term increases and the suppressed direct NLO term decreases by approximately
the same amount with increasing $\xi$. In addition we show also $\dd\sigma/
\dd Q^2$ in the DIS theory, {\it i.e.} without subtraction of any $\ln Q^2$
terms (dashed line). Of course, this cross section must be independent of
$\xi$. This prediction agrees very well with the experimental point, whereas
the result for the subtracted and suppressed theory (full curve) lies
slightly below. We notice, that for $M_{\gamma}=p^{*}_T$ the additional
suppression of DIR$_{\rm IS}$ has only a small effect. It increases
$\dd\sigma/\dd Q^2$ by $5\%$ only.

In order to get an idea about the $M_{\gamma}$ scale dependence of
$\dd\sigma/\dd Q^2$ for the other $Q^2$ bins we have computed this cross
section for two choices of $M_{\gamma}$, namely $M_{\gamma}=p_T^{*}/4$ and
$M_{\gamma}=4p_T^{*}$ corresponding to the lowest and highest $\xi$ in Fig.\
\ref{fig:1} (left). The result for the $\dd\sigma/\dd Q^2$ is shown on the
right side of Fig.\ \ref{fig:1}. We see that the $M_{\gamma}$-dependence in
the considered range decreases with increasing $Q^2$. This is to be expected
since the resolved contribution diminishes with increasing $Q^2$, so that
the NLO corrections to the resolved cross section and the effect of the
evolution of the photon PDF diminish as well. 

\section{Conclusion}
\label{sec:3}

In Summary, we described in this Report a new factorization scheme for
diffractive production of jets in low-$Q^2$ deep inelastic scattering.
By suppressing not only the resolved photon contribution, but also the
unresummed logarithm as well as scheme-dependent finite terms in the NLO
direct initial state correction, factorization scheme and scale invariance
is restored up to higher order effects, while at the same time the cut-off
invariance required in phase space slicing methods is preserved.

For pedagogical reasons, we have chosen in this Report the kinematic region
of finite, but low photon virtuality $Q^2$, which exposes and regularizes a
logarithmic virtual photon initial state singularity. We do, however, not
rely on the finiteness of $Q^2$, but rather separate suppressed and
unsuppressed terms using the hard transverse momentum scale $p_T^*$, so that
our scheme is equally valid for real photoproduction.

The scheme- and scale invariance has been demonstrated numerically using
the kinematics of a recent H1 analysis, differential in $Q^2$ and parton
momentum fraction in the pomeron $z_{\p}$ (not shown). Very good stability 
with respect to scheme- and scale variations and good agreement with the 
experimental data has been found.

\begin{theacknowledgments}
 M.K.\ thanks the working group convenors for the kind invitation and the
 DoE's INT at the University of Washington for kind hospitality and partial
 financial support during preparation of this work.
\end{theacknowledgments}




\begin{thebibliography}{9}

\bibitem{Ingelman:1984ns}
G.~Ingelman and P.~E.~Schlein,
Phys.\ Lett.\ B {\bf 152}, 256 (1985).

\bibitem{Collins:1997sr}
J.~C.~Collins,
Phys.\ Rev.\ D {\bf 57}, 3051 (1998)
[Erratum-ibid.\ D {\bf 61}, 019902 (2000)].

\bibitem{Affolder:2000vb}
T.~Affolder {\it et al.}  [CDF Collaboration],
Phys.\ Rev.\ Lett.\  {\bf 84}, 5043 (2000).

\bibitem{Klasen:2002xb}
M.~Klasen,
Rev.\ Mod.\ Phys.\  {\bf 74}, 1221 (2002).

\bibitem{Klasen:2005dq}
M.~Klasen and G.~Kramer,
DESY 05-095, LPSC 05-053, hep-ph/0506121, submitted to J.\ Phys.\ G.

\bibitem{Klasen:2004tz}
M.~Klasen and G.~Kramer,
contribution to DIS 2004, 
hep-ph/0401202;
%
Eur.\ Phys.\ J.\ C {\bf 38}, 93 (2004).

\bibitem{Klasen:2004ct}
M.~Klasen and G.~Kramer,
Phys.\ Rev.\ Lett.\  {\bf 93}, 232002 (2004).

\bibitem{h1ichep02}
H1 Collaboration,
Abstract 980, contributed to the 31$^{\rm st}$ International Conference
on High Energy Physics (ICHEP 2002), Amsterdam, July 2002.

\bibitem{Chekanov:2004hy}
S.~Chekanov {\it et al.}  [ZEUS Collaboration],
Eur.\ Phys.\ J.\ C {\bf38}, 43 (2004).

\bibitem{Martin:2004xw}
A.~D.~Martin, M.~G.~Ryskin and G.~Watt,
Eur.\ Phys.\ J.\ C {\bf 37}, 285 (2004).

\bibitem{Mozer}
M.~Mozer and R.~Renner for the H1 and ZEUS Collaborations, these
proceedings.

\bibitem{Klasen:1997jm}
M.~Klasen, G.~Kramer and B.~P\"otter,
Eur.\ Phys.\ J.\ C {\bf 1}, 261 (1998).

\bibitem{Potter:1999gg}
B.~P\"otter,
Comput.\ Phys.\ Commun.\  {\bf 133}, 105 (2000).

\bibitem{BKS} D.~B\"odeker, G.~Kramer and S.~G.~ Salesch, Z.\ Phys.\ C
{\bf63}, 471 (1994).
 
\bibitem{Schatzel:2004be}
S.~Sch\"atzel,
hep-ex/0408049,
to appear in the proceedings of the 12$^{\rm th}$ International Workshop on
Deep Inelastic Scattering (DIS 2004), Strbske Pleso, April 2004;
H1 Collaboration,
Abstract 6-0176, contributed to the 32$^{\rm nd}$ International Conference
on High Energy Physics (ICHEP 2004), Beijing, August 2004.

\bibitem{Klasen:1995xe}
M.~Klasen and G.~Kramer,
Phys.\ Lett.\ B {\bf 366}, 385 (1996).

\bibitem{Adloff:2000qi}
C.~Adloff {\it et al.}  [H1 Collaboration],
Eur.\ Phys.\ J.\ C {\bf 20}, 29 (2001).

\bibitem{Schuler:1996fc}
G.~A.~Schuler and T.~Sj\"ostrand,
Phys.\ Lett.\ B {\bf 376}, 193 (1996).

\end{thebibliography}
\end{document}